\documentclass[a4paper,aps,prd,onecolumn,preprintnumbers,showpacs,nofootinbib]{revtex4}
\usepackage{amsmath,amssymb,graphics,epsfig,subfigure}
\usepackage{color}

\begin{document}
\newcommand {\nn} {\nonumber}
\renewcommand{\baselinestretch}{1.3}

\title{On the critical phenomena and thermodynamic geometry of charged Gauss-Bonnet AdS black hole}

\author{Shao-Wen Wei \footnote{weishw@lzu.edu.cn},
        Yu-Xiao Liu \footnote{liuyx@lzu.edu.cn, corresponding author}}

\affiliation{Institute of Theoretical Physics, Lanzhou University, Lanzhou 730000, People's Republic of China}

\begin{abstract}
In this paper, we study the phase structure and equilibrium state space geometry of charged topological Gauss-Bonnet black holes in $d$-dimensional anti-de Sitter spacetime. Several critical points are obtained in the canonical ensemble, and the critical phenomena and critical exponents near them are examined. We find that the phase structures and critical phenomena drastically depend on the cosmological constant $\Lambda$ and dimensionality $d$. The result also shows that there exists an analogy between the black hole and the van der Waals liquid gas system. Moreover, we explore the phase transition and possible property of the microstructure using the state space geometry. It is found that the Ruppeiner curvature diverges exactly at the points where the heat capacity at constant charge of the black hole diverges. This black hole is also found to be a multiple system, i.e., it is similar to the ideal gas of fermions in some range of the parameters, while to the ideal gas of bosons in another range.
\end{abstract}



\maketitle

\section{Introduction}
\label{secIntroduction}

Since the discovery of the four laws of black hole mechanics, black holes are widely believed to be thermodynamic objects assigned standard thermodynamic variables such as the temperature and entropy \cite{Bardeen,Bekenstein}. Even though the exact statistical description of black hole microstates is still unclear, the study of the thermodynamic properties of black holes has been a fascinating topic. These black holes are found to possess the interesting phase transition, for example the Hawking-Page phase transition \cite{Hawking}, and critical phenomena as seen in normal thermodynamic systems.

Motivated by the AdS/CFT correspondence \cite{Maldacena}, where the transitions have been related with the holographic superconductivity \cite{Gubser,Hartnoll}, the subject that the phase transitions of black holes in asymptotically anti de-Sitter (AdS) spacetime, has received considerable attention \cite{Wu00prd,Cvetic,CveticGubser,Sahay2,Sahay,Sahay3,Banerjee10prd,Chamblin,
Jiang,ZhangLi,Lalaaaa,Kastor}. The underlying microscopic statistical interaction of the black holes is also expected to be understood via the study of the gauge theory living on the boundary in the gauge/gravity duality.

Recently, the authors of \cite{Niu} studied the phase transitions and critical phenomena for the Reissner-Nordstrom (RN) black hole in $(n+1)$-dimensional AdS spacetime; they found that the black hole systems is similar to a van der Waals liquid gas system. The critical exponents near a van der Waals-like critical point were found to be the same as the liquid gas system. Especially, near the critical point, the $Q$-$\Phi$ diagram was presented, sharing the same shape as that of $P$-$V$ diagram for a van der Waals liquid gas system, which strongly suggests a remarkable analogy between these two thermodynamic systems. Almost at the same time, the similar critical behavior was found in the Kerr-AdS black hole \cite{Tsai}. Different from the RN-AdS black hole, there are two van der Waals-like critical points identified at different temperatures. The critical phenomena of Born-Infeld AdS black holes and higher curvature charged AdS black holes were investigated in \cite{Banerjee,Banerjee3,Lala}, where the critical exponents were also examined. Treating the cosmological constant as a thermodynamic pressure and its conjugate quantity as a thermodynamic volume, the critical behavior and $P$-$V$ criticality of a charged AdS black hole is reexamined \cite{Kubiznak,Gunasekaran,Dolan}. And a complete analogy between the black hole and the liquid-gas system was established \cite{Kubiznak}.

In a separate context, a geometrical perspective of thermodynamics has been a subject of great interest for several decades. Following the development of the extrinsic geometrical perspective of thermodynamics \cite{Tisza,Callen}, Weinhold \cite{Weinhold} proposed an intrinsic geometrical perspective defined in the energy representation on the space of the equilibrium thermodynamic states of a system. However, this geometry cannot be well interpreted from the physical view. On the other hand, Ruppeiner \cite{Ruppeiner} constructed an intrinsic Riemannian geometry in the entropy representation, which is found to be related to the probability measuring the thermodynamic fluctuations connecting two equilibrium states. It is also suggested that the scalar curvature produced from this geometry contains the information of the phase transitions and critical phenomena, and it shares the same divergent behavior as the heat capacity of that thermodynamic system.

Also, Ruppeiner pointed out that the sign of the curvature may be an indicator of the interactions among the microscopic structure of the thermodynamic system \cite{Ruppeiner2}. For example, the curvature is positive for the repulsive interaction, and negative for the attractive interaction, which correspond to the ideal gas of fermions and bosons, respectively \cite{Janyszek,Oshima}. As we know, there is attractive interaction among the microscopic particles for the van der Waals system, which implies a negative curvature. And the result is true, the curvature for that thermodynamic system is negative and goes to negative infinity at the critical point. It is suggested that the negative divergent behavior of the curvature is related to the Bose-Einstein condensation, and the positive divergent behavior is related to the Pauli's exclusion principle.

Among the gravity theories with higher derivative curvature terms, the Gauss-Bonnet (GB) gravity has some special features and gives rise
to some interesting effects on the thermodynamics of black holes in AdS space \cite{Cai02,Odintsov,Banados,Clunan,Kofinas,Kim,Park,Wang,Astefanesei}. The phase structure of a GB-AdS black hole was briefly studied in \cite{Cai02,Dey}. And in the grand canonical ensemble, the local and global thermal phase structure of a charged asymptotically AdS black hole with both GB and quartic field strength corrections were thoroughly researched \cite{Anninos}.

In this paper, we will study the phase transition and critical behavior of charged topological GB black holes in $d$-dimensional AdS spacetime. Motivated by the idea that the cosmological constant $\Lambda$ can be regarded as a thermodynamic variable \cite{Kubiznak}, we leave it free to vary and study the phase structures and critical phenomena for the GB-AdS black hole in the canonical ensemble. The result strongly suggests that the phase structures and critical phenomena drastically depend on $\Lambda$ and $d$. Then we examine the thermodynamic analogy between a charged GB-AdS black hole and a van der Waals liquid gas system. The $Q$-$\Phi$ criticality is investigated, which is analogous to the $P$-$V$ criticality of a van der Waals system. The critical exponents are calculated and found to be the same as that of van der Waals liquid gas system. Moreover, we explore the phase transitions and possible property among the microstructure of a charged GB-AdS black hole from the viewpoint of geometrical perspective of equilibrium thermodynamics. We find the scalar curvature diverges precisely at the divergent points where the heat capacity diverges. This black hole system is also found to be a multiple system, i.e., it is similar to the ideal gas of fermions in some range of the parameters, while similar to the ideal gas of bosons in another range. And both the phenomena of Bose-Einstein condensation and Pauli's exclusion are expected to be observed in such thermodynamic system.

This paper is organized in the following manner. In Sec. \ref{Thermodynamics}, we review some thermodynamic properties of the charged GB-AdS black hole. For different values of $d$ and $\Lambda$, the phase structures and critical behaviors are well studied in Sec. \ref{Critical}.
In Sec. \ref{exponent}, the critical exponents and scaling symmetry are examined. In Sec. \ref{curvature}, we investigate the phase transitions and possible property among the microstructure of a charged GB-AdS black hole from the viewpoint of geometrical perspective of equilibrium thermodynamics. Sec. \ref{Conclusion} is devoted to the summary and conclusion of our results.

\section{Review of the thermodynamics}
\label{Thermodynamics}

The action of the $d$-dimensional spacetime in Einstein-Gauss-Bonnet gravity with a negative cosmological constant $\Lambda=-(d-1)(d-2)/2l^{2}$ has the form
\begin{eqnarray}
 S=\frac{1}{16\pi G_{d}}\int d^{d}x\sqrt{-g}(\mathcal{R}-2\Lambda
                         +\alpha_{GB}\mathcal{L}_{GB}
                         -\mathcal{L}_{matter}),
\end{eqnarray}
where $\alpha_{GB}$ is the GB coupling constant with dimension $(length)^{2}$ and it is regarded as the inverse string tension with positive value. The GB Lagrangian $\mathcal{L}_{GB}$ and the electromagnetic Lagrangian $\mathcal{L}_{matter}$ are
\begin{eqnarray}
&&\mathcal{L}_{GB}=\mathcal{R}_{\mu\nu\gamma\delta}\mathcal{R}^{\mu\nu\gamma\delta}
                    -4\mathcal{R}_{\mu\nu}\mathcal{R}^{\mu\nu}+\mathcal{R}^{2},\\
&&\mathcal{L}_{matter}=4\pi
G_{d}\mathcal{F}_{\mu\nu}\mathcal{F}^{\mu\nu}.
\end{eqnarray}
where $\mathcal{R}$, $\mathcal{R}_{\mu\nu}$, and
$\mathcal{R}_{\mu\nu\gamma\delta}$ are the Ricci scalar, the Ricci tensor, and the Riemann tensor, respectively. The electromagnetic tensor field
$\mathcal{F}_{\mu\nu}=\partial_{\mu}\mathcal{A}_{\nu}-\partial_{\nu}\mathcal{A}_{\mu}$
with $\mathcal{A}_{\mu}$ the vector potential.

The $d$-dimensional static charged GB-AdS black hole solution for the action is described by (we use the units $G_{d}=\hbar=c=k_{B}=1$)
\begin{eqnarray}
 ds^{2}=-f(r)dt^{2}+f^{-1}(r)dr^{2}+r^{2}f_{ij}dx^{i}dx^{j},\label{metric}
\end{eqnarray}
where $f_{ij}dx^{i}dx^{j}$ is the line element of a
$(d-2)$-dimensional hypersurface with constant curvature
$(d-2)(d-3)k$ and volume $\Sigma_{k}$ ($k$ is a horizon curvature constant), having an explicit form
\begin{eqnarray}
 f_{ij}dx^{i}dx^{j}=\left\{
                    \begin{array}{ll}
                      d\theta^{2}+\sin^{2}\theta d\phi^{2}+\cos^{2}\theta d\Omega^{2}_{d-4}, & \hbox{$k$=1\;\;(spherical symmetry),} \\
                      d\theta^{2}+d\phi^{2}+dx^{2}_{d-4},
                   & \hbox{$k$=0\;\;(planar symmetry),} \\
                      d\theta^{2}+\sinh^{2}\theta d\phi^{2}+\cosh^{2}\theta d\Omega^{2}_{d-4}, & \hbox{$k$=-1\;(hyperbolic symmetry).}
                    \end{array}
                  \right.
\end{eqnarray}
The metric function is \cite{Odintsov,Wiltshir}
\begin{eqnarray}
 f(r)=k+\frac{r^{2}}{2\alpha}\bigg(1-\sqrt{1-\frac{4\alpha}{l^{2}}}
           \sqrt{1+\frac{m}{r^{d-1}}-\frac{q^{2}}{r^{2d-4}}}\bigg),
\end{eqnarray}
where $\alpha=(d-3)(d-4)\alpha_{GB}$. The parameter $m$ and $q$ are corresponded to the gravitational mass $M$ and charge $Q$, respectively, as
\begin{eqnarray}
 M&=&\frac{(d-2)\Sigma_{k}(1-4\alpha/l^{2})}{64\pi\alpha}m,\\
 Q^{2}&=&\frac{\pi(d-2)(d-3)(1-4\alpha/l^{2})}{2\alpha}q^{2}.
\end{eqnarray}
For this black hole solution, the horizon radius $r_{h}$ is determined by the equation $f(r)=0$. Then the gravitational mass can be reexpressed as
\begin{eqnarray}
 M=\frac{(d-1)Q^{2}r_{h}^{8}+2\pi r_{h}^{2d}(d-3)
             \bigg((d^{2}-3d+2)(kr_{h}^{2}+k^{2}\alpha)-2\Lambda r_{h}^{4}\bigg)}
      {32\pi^{2}(d^{2}-4d+3)r_{h}^{d+5}}\Sigma_{k}.\label{mass}
\end{eqnarray}
For the metric (\ref{metric}) to be free from conical singularity at $r=r_{h}$, we get the temperature $T$ of the black hole
\begin{eqnarray}
 T=\frac{-Q^{2}r_{h}^{8}+2\pi r_{h}^{2d}
            \bigg((d-2)k((d-3)r_{h}^{2}+(d-5)k\alpha)
                  -2\Lambda r_{h}^{4}\bigg)}
            {8\pi^{2}(d-2)r_{h}^{2d+1}(2k\alpha+r_{h}^{2})}.\label{tempature}
\end{eqnarray}
As we known, an extremal black hole has a vanishing temperature $T$. So, solving $T=0$, one can obtain
\begin{eqnarray}
 Q_{e}^{2}=2 \pi r_{h}^{2 d-8}\left((d-2) k \left((d-5)k\alpha
    +(d-3)r_{h}^2\right)-2r_{h}^4 \Lambda \right).
\end{eqnarray}
The range of $T<0$ represents the non-black hole case, i.e., naked singularity. So, we will only focus on the range of $T>0$ in this paper. The potential difference between the horizon and infinity is
\begin{eqnarray}
 \Phi=\bigg(\frac{\partial M}{\partial Q}\bigg)_{S}
     =\frac{Q \Sigma_{k}}{16\pi^{2}(d-3)r_{h}^{d-3}}.\label{phi}
\end{eqnarray}
By employing the first law of black hole thermodynamics, we obtain the entropy for the black hole:
\begin{eqnarray}
 S=\int\frac{\partial_{r_{h}}M(r_{h},Q)}{T}dr_{h}
  =\frac{\Sigma_{k}}{4 r_{h}^{2-d}}
     \bigg(1+\frac{2k\alpha(d-2)}{(d-4)r_{h}^{2}}\bigg).\label{entropy}
\end{eqnarray}
The Gibbs free energy defined as $G=M-TS-Q\Phi$ is
\begin{eqnarray}
 G=&\frac{\Sigma_{k}}{32}
   \bigg(\frac{2 \pi  (d-3) r_{h}^{d-5} \left(\left(d^2-3 d+2\right) k^2
   \alpha +\left(d^2-3 d+2\right) k r_{h}^2-2 r_{h}^4 \Lambda
   \right)+(d-1) Q^2 r_{h}^{3-d}}{\pi^2 \left(d^2-4 d+3\right)}\nonumber\\
  &-8 T r_{h}^{d-2} \left(\frac{2 (d-2)k \alpha }
    {(d-4)r_{h}^2}+1\right)-\frac{2 Q^2 r_{h}^{3-d}}{\pi ^2 (d-3)}\bigg).
\end{eqnarray}
Another important thermodynamical quantity measuring the stability against small perturbation is the heat capacity $C_{Q}$ at constant charge, defined as
\begin{eqnarray}
 C_{Q}=T\left(\frac{\partial S}{\partial T}\right)_{Q}
     =-\frac{2\pi^{2}r_{h}^{3d}\Sigma_{k}(d-2)(r_{h}^{2}+2k\alpha)^{3}}
       {\Delta}T,\label{heatcapacity}
\end{eqnarray}
where
\begin{eqnarray}
 \Delta=&2\pi r_{h}^{2d+3}\bigg((d-2)k\left(2(d-5)k^2\alpha^2+(d-9)k
   r_{h}^2\alpha+(d-3)r_{h}^4\right)+2r_{h}^4\Lambda\left(6k\alpha+r_{h}^2\right)
   \bigg)\nonumber\\
   &-Q^2r_{h}^{11}\left(2(2d-7)k\alpha+(2d-5)r_{h}^2\right).
\end{eqnarray}
From (\ref{heatcapacity}), we see that the heat capacity $C_{Q}$ vanishes for an extremal black hole ($T=0$). While, for a nonextremal black hole, $C_{Q}$ is nonvanishing. In general, the local stability is determined by the heat capacity, which states that the positive heat capacity can guarantee a stable black hole to exist, while the negative one will make the black hole disappear when a small perturbation is included within. Moreover, solving $\Delta=0$, we get the divergent point of the heat capacity, which is
\begin{eqnarray}
 Q_{d}^{2}=2\pi\frac{k(d-2)\big((d-3)r_{h}^{4}+(d-9)k\alpha r_{h}^{2}+2(d-5)k^{2}\alpha^{2}\big)+2r_{h}^{4}(r_{h}^{2}+6k\alpha)\Lambda}
   {r_{h}^{8-2d}\left((2d-5)r_{h}^{2}+2k\alpha(2d-7)\right)}.\label{Qccondition}
\end{eqnarray}
The parameter $K_{T}$, which is related to the isothermal compressibility is
\begin{eqnarray}
 K_{T}=Q^{-1}\bigg(\frac{\partial\Phi}{\partial Q}\bigg)_{T}
           =-\bigg[Q\left(\frac{\partial\Phi}{\partial T}\right)_{Q}
             \left(\frac{\partial T}{\partial Q}\right)_{\Phi}\bigg]^{-1},
\end{eqnarray}
where the thermodynamic identity $\left(\frac{\partial \Phi}{\partial T}\right)_{Q}\left(\frac{\partial T}{\partial Q}\right)_{\Phi}\left(\frac{\partial Q}{\partial \Phi}\right)_{T}=-1$ is used. Using (\ref{tempature}) and (\ref{phi}), we get
\begin{eqnarray}
 K_{T}^{-1}=-\frac{Q\Sigma_{k}(r_{h}^{2}-2k\alpha)(Q^{2}-\Theta)}
         {16(d-3)\pi^{2}r_{h}^{d-14}\Delta},\label{kt}
\end{eqnarray}
with
\begin{eqnarray}
 \Theta=\frac{(2d-5)r_{h}^{2}+2k\alpha (2d-7)}{r_{h}^2-2 k\alpha}Q_{d}^{2}.
\end{eqnarray}
From (\ref{kt}), one easily obtains the result that $K_{T}^{-1}$ diverges exactly at the points where the heat capacity $C_{Q}$ diverges and vanishes at $Q^{2}=\Theta$. A similar result also holds for the Kerr-Newman black hole \cite{Lousto}, and Born-Infeld AdS black hole \cite{Banerjee}.

It is also clear that, thermodynamic properties of the black hole closely depend on the hypersurface parameter $k$ and the dimension of the spacetime $d$. So in the following, we will discuss these cases, respectively. And without loss of generality, we take the choice $\Sigma_{k}=1$ and $\alpha=1$.

\section{Phase structure and $Q$-$\Phi$ criticality}
\label{Critical}

In this section, we will study the phase structure for the black hole of different $k$ and $d$ in the canonical ensemble with the charge $Q$ fixed.

\subsection{$k=1$}

In this case, the divergence condition (\ref{Qccondition}) for the heat capacity $C_{Q}$ will reduce to
\begin{eqnarray}
 Q_{d}^{2}=2\pi r_{h}^{2d-8}\frac{(d-2)\big[(d-3)r_{h}^{4}+(d-9) r_{h}^{2}+2(d-5)\big]-2r_{h}^{4}(r_{h}^{2}+6)}
   {(2d-5)r_{h}^{2}+2(2d-7)}.
\end{eqnarray}

For $d=5$, this condition becomes
\begin{eqnarray}
 Q_{d}^{2}=-4\pi r_{h}^{4}\frac{r_{h}^{4}+3r_{h}^{2}+6}{5r_{h}^{2}+6}.
\end{eqnarray}
From above, we find $Q_{d}^{2}$ is negative for an arbitrary value of $r_{h}$. So, the heat capacity $C_{Q}$ of a five-dimensional black hole is always regular, which implies that there is no phase transition points related to the divergent points of $C_{Q}$.

\begin{figure*}
\centerline{\subfigure[]{\label{Cqk1a}
\includegraphics[width=8cm,height=6cm]{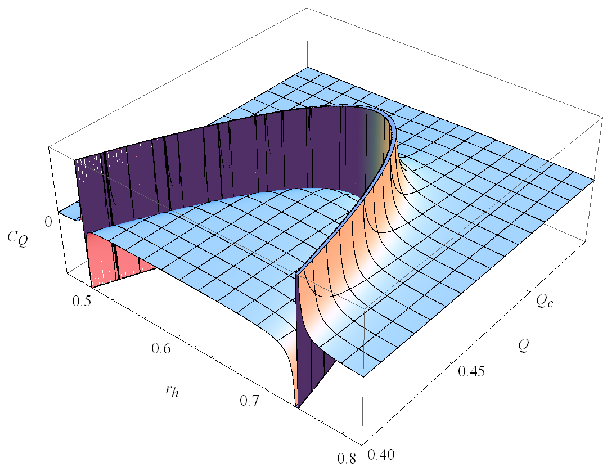}}
\subfigure[]{\label{Cqk1b}
\includegraphics[width=8cm,height=6cm]{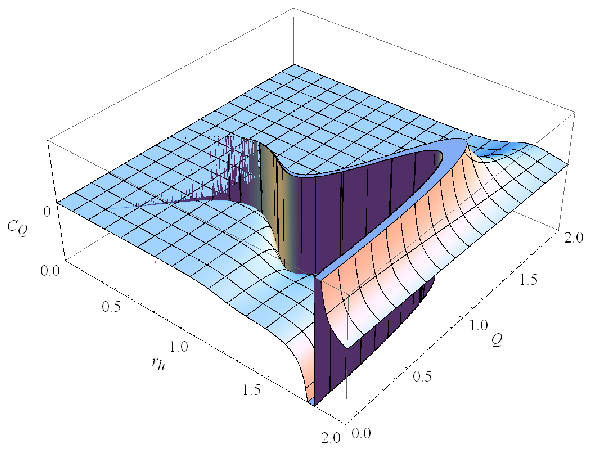}}}
\caption{Heat capacity $C_{Q}$ with $r_{h}$ and $Q$ at $k=1$. (a) for $d=6$, $\Lambda=-1$, and (b) for $d=6$, $\Lambda=-0.49$.} \label{PCqk1}
\end{figure*}

For $d=6$,  the condition reduces to
\begin{eqnarray}
 Q_{d}^{2}=4\pi r_{h}^{4}\frac{r_{h}^{6}6r_{h}^{4}(1+\Lambda)
           -6r_{h}^{2}+4}{7r_{h}^{2}+10}.
\end{eqnarray}
In this case, it is clear that the divergent behavior of the heat capacity depends crucially on the cosmological constant $\Lambda$. After a detail examination, we find that the heat capacity $C_{Q}$ has two different behaviors for $\Lambda\in(-\infty,-0.5)\cup(-0.4725,0)$, and $\Lambda\in(-0.5,-0.4725)$. In Fig. \ref{PCqk1}, we plot the heat capacity $C_{Q}$ for the $\Lambda=-1$ and $-0.49$, respectively.

\begin{figure*}
\centerline{\subfigure[]{\label{Qdk1a}
\includegraphics[width=8cm,height=6cm]{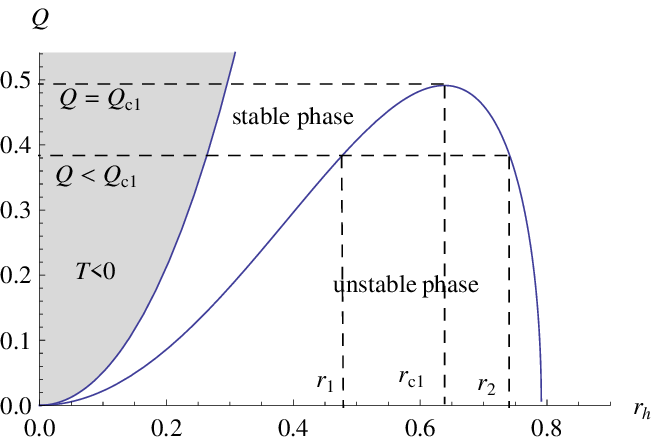}}
\subfigure[]{\label{Qdk1b}
\includegraphics[width=8cm,height=6cm]{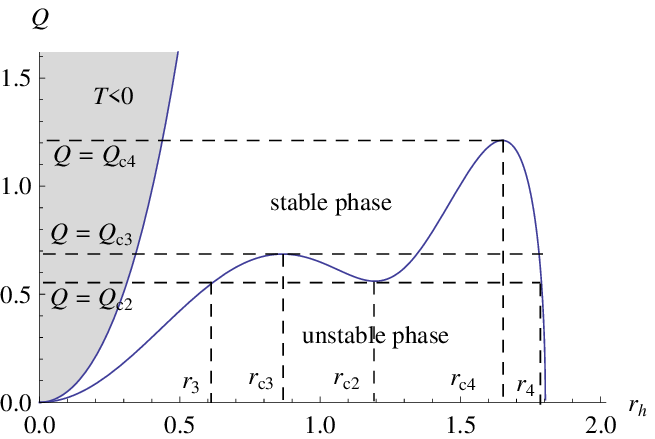}}}
\caption{The divergent behavior of the heat capacity as a function of $r_{h}$. (a) for $d=6$, $\Lambda=-1$. (b) for $d=6$, $\Lambda=-0.49$. The range in shadow has a temperature $T<0$.} \label{PQdk1}
\end{figure*}

Case 1: $\Lambda\in(-\infty,-0.5)\cup(-0.4725,0)$. From Fig. \ref{Cqk1a}, we can see that, for small value of charge $Q$, with the increase of $r_{h}$, the heat capacity first goes to positive infinity at where we denote as $r_{1}$. Then it increases from negative infinity to a finite negative value and back to negative infinity at $r_{h}=r_{2}$. And at last, it decreases from positive infinity to a finite positive value and then monotonically increases to infinity at $r_{h}=\infty$. From the above analysis, it is quite evident that the heat capacity $C_{Q}$ is positive for $r_{h}<r_{1}$ and $r_{h}>r_{2}$, while it is negative for $r_{1}<r_{h}<r_{2}$. Moreover, $C_{Q}$ suffers discontinuities at $r_{h}=r_{1},r_{2}$, which correspond to the phase transition points of the charged GB-AdS black holes. The point $r_{1}$ corresponds to the transition that a small stable black hole with $C_{Q}>0$ to an intermediate unstable one with $C_{Q}<0$. And the point $r_{2}$ corresponds to an intermediate unstable black hole to a large stable one. On the other hand, for the large value of charge $Q$, the divergent behavior vanishes and the black hole is in a stable phase. To better understand the divergent behaviors of the heat capacity, we plot the divergent point in the $(Q,\;r_{h})$ plane in Fig. \ref{Qdk1a}. The shadow region in this figure has a negative temperature, which represents a non-black hole case, and we will not discuss it here. From it, we see that, for $Q<Q_{c1}$, there exist two divergent points $r_{1}$ and $r_{2}$; for $Q=Q_{c1}$, the two divergent points coincide with each other at $r_{h}=r_{c1}$; and for $Q>Q_{c1}$, the divergent point disappears. Therefore, $Q_{c1}$ corresponds to a critical phase transition point and it is a local maxima along this divergent curve.

Case 2: $\Lambda\in(-0.5,-0.4725)$. In this case, $C_{Q}$ has a complex structure; see Fig. \ref{Cqk1b}. On one hand, like the case 1, the small and larger black holes are in thermally stable phase, while the intermediate black hole is unstable. On the other hand, from the Fig. \ref{Qdk1b} describing the divergent behavior in the $(Q,\;r_{h})$ plane, one finds that there exist two local maxima located at $r_{h}=r_{c3}$, $r_{c4}$ and one local minimum located at $r_{h}=r_{c2}$. For a small charge, there are two divergent points. When $Q=Q_{c2}$, we have three divergent points located at $r_{h}=r_{3}, r_{2c}$, and $r_{4}$, respectively. When $Q$ further increases, the divergent point at $r_{c2}$ is split into two. One coincides with $r_{c3}$ at $Q=Q_{c3}$, and the other one meets $r_{c4}$ at $Q=Q_{c4}$. For $Q>Q_{c4}$, the divergent point disappears. Moreover, we have the following relations, $Q_{c2}<Q_{c3}<Q_{c4}$, and $r_{c3}<r_{c2}<r_{c4}$. It is also worth noting that, $r_{c3}$ will meet $r_{c2}$ at $\Lambda=-0.4725$. Here $r_{h}=r_{c2}$, $r_{c3}$, and $r_{c4}$ are corresponded to the different critical phase transition points. In order to study the $Q$-$\Phi$ criticality, we list the numerical values of the thermodynamic quantities at the critical points in Table. \ref{criticalparameters}. This critical behavior is very interesting and different from the RN-AdS \cite{Niu}, and BI-AdS cases \cite{Banerjee,Banerjee3}, where only one critical point located at the maximum was found. And it is also different from the Kerr-AdS black hole \cite{Tsai}. Here we have three critical points, and two of them have the same temperature.


For $d\geq 7$, the heat capacity and its divergent behavior are almost the same as that shown in Figs. \ref{Cqk1a} and \ref{Qdk1a}, for an arbitrary value of the cosmological constant $\Lambda$. We also list the critical parameters $r_{c}$, $Q_{c}$, $\Phi_{c}$, $T_{c}$ and radius $r_{e}$ of the extremal black hole in Table \ref{criticalparameters}. From it, we can see that the critical parameters $r_{c}$, $Q_{c}$, and $T_{c}$ increase with the dimensionality $d$. $r_{c}$ and $T_{c}$ vary slowly with $d$, while $Q_{c}$ increases quickly when $d$ increases. Compared with them, the critical potential $\Phi_{c}$ has a complicated behavior, i.e., it first decreases, and then increases, and at last decreases with $d$. It is also worth noting that, for different $d$, we always have $r_{e}<r_{c}$, which implies that these critical phenomena correspond to the nonextremal black hole, and it is a nontrivial phase transition.

\begin{table}[h]
\begin{center}
\caption{The values of these critical parameters $r_{c}$, $Q_{c}$, $\Phi_{c}$, $T_{c}$ and the extremal black hole radius $r_{e}$ for different dimensionality $d$, and cosmological constant $\Lambda$.}\label{criticalparameters}
\begin{tabular}{c c c c c c c}
  \hline
  \hline
   $d$&$\Lambda$ &$r_{c}$ & $Q_{c}$ & $\Phi_{c}$ & $T_{c}$ & $r_{e}$ \\
  \hline
    6 & -0.49 &0.8664 & 0.6857 & 0.0022 & 0.1121 & 0.3426 \\
    6 & -0.49 &1.1952 & 0.5604 & 0.0007 & 0.1122 & 0.3132 \\
    6 & -0.49 &1.6489 & 1.2116 & 0.0006 & 0.1121 & 0.4376 \\\hline
    6 & -1 &0.6382 & 0.4914 & 0.0040 & 0.1164 & 0.2952 \\
    7 & -1&1.6425 & 6.2129 & 0.0014 & 0.1613 & 0.7991 \\
    8 & -1&2.7273 & 177.87 & 0.0015 & 0.1804 & 1.6143 \\
    9 & -1&3.6039 & 2990.1 & 0.0014 & 0.1909 & 2.3218 \\
   10 & -1&4.4157 & 48147  & 0.0013 & 0.1978 & 2.9965 \\
  \hline\hline
\end{tabular}
\end{center}
\end{table}

\begin{figure*}
\subfigure[]{\label{PQphita}
\includegraphics[width=8cm,height=6cm]{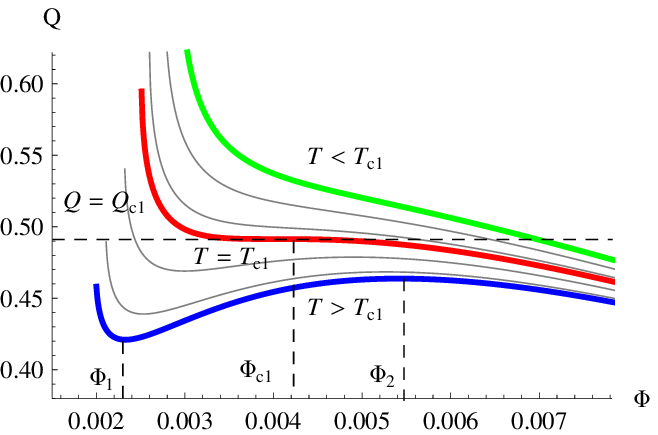}}
\subfigure[]{\label{PQphitd}
\includegraphics[width=8cm,height=6cm]{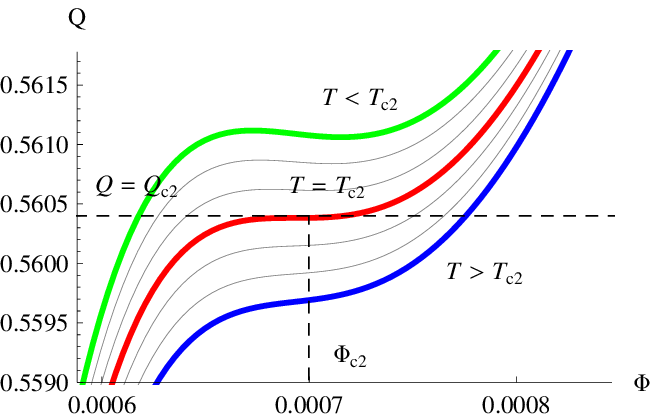}}\\
\subfigure[]{\label{PQphitb}
\includegraphics[width=8cm,height=6cm]{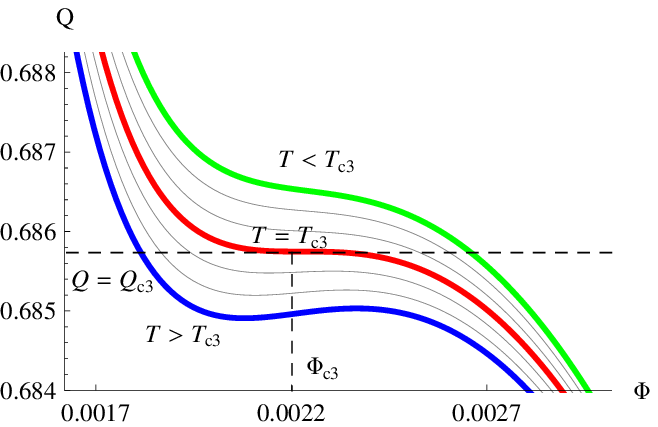}}
\subfigure[]{\label{PQphitc}
\includegraphics[width=8cm,height=6cm]{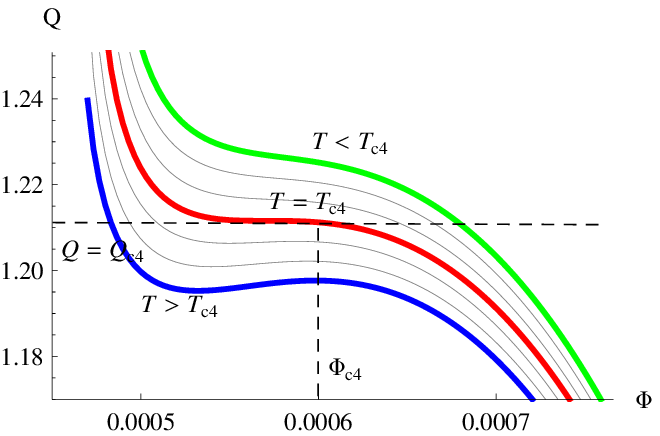}}
\caption{$Q$-$\Phi$ diagram (isotherm) near the four critical points located at $r_{c1}$, $r_{c2}$, $r_{c3}$, and $r_{c4}$, respectively.} \label{PQphit}
\end{figure*}

\begin{figure*}
\centerline{
\includegraphics[width=8cm,height=6cm]{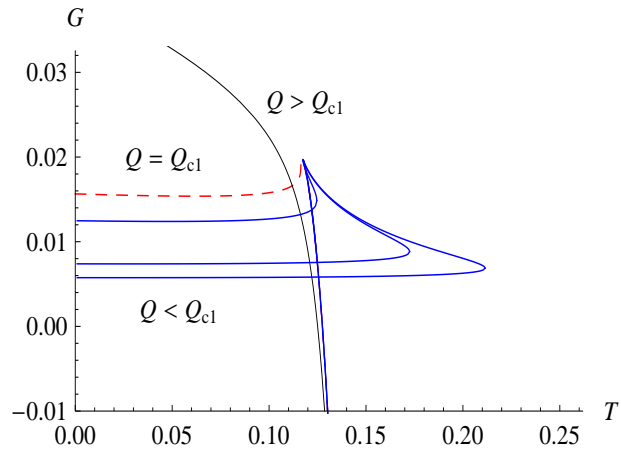}}
\caption{Gibbs free energy as a function of the temperature $T$ for $\Lambda=-1$. The `swallow tail' behavior appears when $Q<Q_{c1}$.} \label{PGT}
\end{figure*}

Now let us turn to the $Q$-$\Phi$ criticality of the black hole. With the potential (\ref{phi}), the equation of state of this case for any $d$ reads
\begin{eqnarray}
 T=\frac{-Q^{2}\Phi_{0}^{8-2d}+2\pi\left(10-7d+d^{2}
          +(d-2)(d-3)\Phi_{0}^{2}+2\Phi_{0}^{4}\right)}
          {8\pi^{2}(d-2)(2+\Phi_{0}^{2})\Phi_{0}},\label{Tphiq}
\end{eqnarray}
where $\Phi_{0}=\left(16\pi^{2}(d-3)\frac{\Phi}{Q}\right)^{1/(3-d)}$.

For $d=6$, it is clear that the phase structure is different for $\Lambda\in(-\infty,-0.5)\cup(-0.4725,0)$ and $\Lambda\in(-0.5,-0.4725)$. So, the $Q$-$\Phi$ diagram will also behave different. At first, we consider the case $\Lambda=-1$. As presented in Fig. \ref{Qdk1a}, when the charge of a GB-AdS black hole reaches the critical value $Q_{c1}$, the two divergent points located at $r_{1}$ and $r_{2}$ degenerate into a single one located at $r_{c}$. The thermally unstable phase of a GB-AdS black hole will disappear with a further increase of the charge. From (\ref{Tphiq}), we can obtain the isotherm of a GB-AdS black hole. The result is shown in Fig. \ref{PQphita}. It is obvious that this $Q$-$\Phi$ diagram is very similar to the $P$-$V$ diagram of a van der Waals liquid gas system. An isotherm of a GB-AdS black hole with $T>T_{c1}$ has a local maximum and minimum located at $\Phi_{2}$ and $\Phi_{1}$, respectively. Along the segment of the isotherm between the two points, we have $\left(\frac{\partial Q}{\partial \Phi}\right)_{T}>0$, which implies that the black holes are in a thermally unstable phase. While, for the $\left(\frac{\partial Q}{\partial \Phi}\right)_{T}<0$ region, these black holes are in a thermally stable phase corresponding to the small and large black holes. When the critical temperature $T_{c}$ is reached, the shape of the isotherm will undergo significant change. The two points located at $\Phi_{1}$ and $\Phi_{2}$ merge into a single one located at $\Phi_{c1}$. It is worth noting that the critical point at $\Phi_{c1}$ coincides with that located at $r_{c1}$ on the critical isocharge curve with $Q=Q_{c}$ shown in Fig. \ref{Qdk1a}. When $T>T_{c}$, the isotherm will have no local maxima or minima, and it monotonously decreases with $\Phi$, which means that the thermally unstable phase disappears.

Here, let us turn to the second case, where $\Lambda\in(-0.5,-0.4725)$. We take $\Lambda=-0.49$ as an example. There exists a more richer phase structure than the $\Lambda=-1$ case. As shown in Fig. \ref{Cqk1b}, two local maxima and one local minimum appear, and the critical values of these thermodynamic quantities are listed in Table \ref{criticalparameters}. One thing worth noting is that the critical values of the temperature $T_{c3}=T_{c4}$, which means the critical points located at $r_{c3}$ and $r_{c4}$ lie on the same isotherm of a GB-AdS black hole. The $Q$-$\Phi$ criticality of this case is shown in Figs. \ref{PQphitb} and \ref{PQphitc}. It is clear that they are the similar to that of Fig. \ref{PQphita} describing the $\Lambda=-1$ case. These are for the two local maxima located at $r_{c3}$ and $r_{c4}$, respectively. When $T>T_{c3}$ $(T_{c4})$, there exists thermally unstable phase, or the unstable phase will disappear. For the local minimum one, the $Q$-$\Phi$ diagram is shown in Fig. \ref{PQphitd}. Its behavior is very different from the local maximum one. For $T<T_{c2}$, thermally stable phase can be found, and for $T>T_{c2}$, only the unstable phase is allowed. Combining with these figures, we can get the range $T_{c3}(=T_{c4})<T<T_{c2}$, where the thermally unstable phase can exist. The Gibbs free energy is plotted in Fig. \ref{PGT} for $\Lambda=-1$. Since the $G$ curve demonstrates the characteristic ``swallow tail" behavior, there is a first order transition in this black hole system. And for other critical points, this swallow tail behavior can also be found.

For $d\geq 7$, its phase structure is almost the same as that described in Figs. \ref{Cqk1a} and \ref{Qdk1a}. For any value of $\Lambda$, only one local maximum appears. Therefore, the shape of its $Q$-$\Phi$ diagram is similar as the one shown in Fig. \ref{PQphita}.

\subsection{$k=0$}

For this planar symmetry case, the divergence condition (\ref{Qccondition}) for the heat capacity $C_{Q}$ can be expressed in a compact form
\begin{eqnarray}
 Q_{d}^{2}=-\frac{4\pi r_{h}^{2d-4}}{2d-5}.
\end{eqnarray}
Since $d\geq 5$, $Q_{d}^{2}$ is negative. Thus the heat capacity $C_{Q}$ in this case is regular for any values of these parameters, which implies that there exists no phase transition. The $Q$-$\Phi$ diagram is a trivial one. So, we skip this case.

\subsection{$k=-1$}

\begin{figure*}
\centerline{\subfigure[]{\label{PCqk-1a}
\includegraphics[width=8cm,height=6cm]{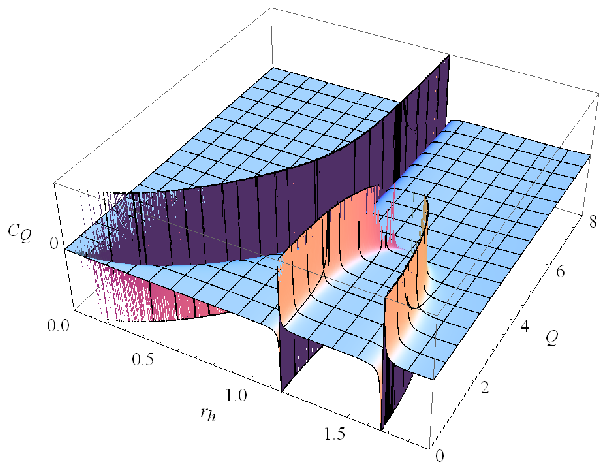}}
\subfigure[]{\label{PCqk-1b}
\includegraphics[width=8cm,height=6cm]{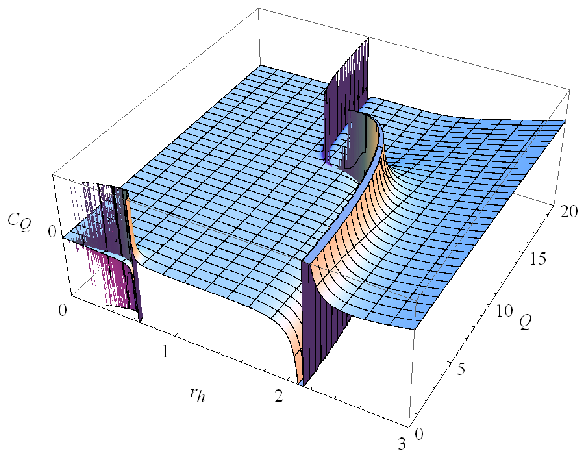}}}
\caption{Heat capacity $C_{Q}$ with $r_{h}$ and $Q$ at $k=-1$. (a) for small $\Lambda$, and (b) for large $\Lambda$.} \label{PCqk}
\end{figure*}

\begin{figure*}
\subfigure[]{\label{PQdk-1a}
\includegraphics[width=8cm,height=6cm]{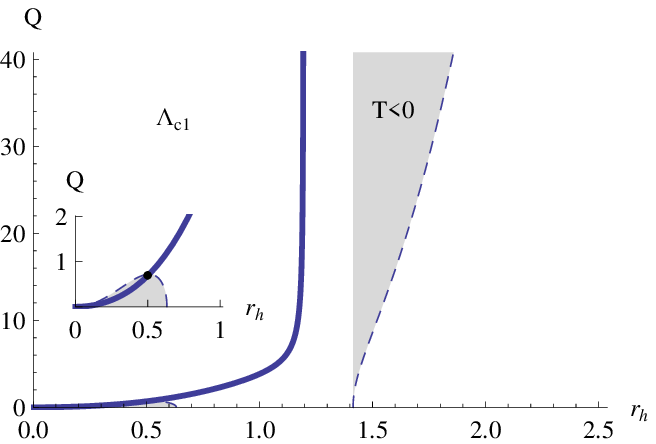}}
\subfigure[]{\label{PQdk-1b}
\includegraphics[width=8cm,height=6cm]{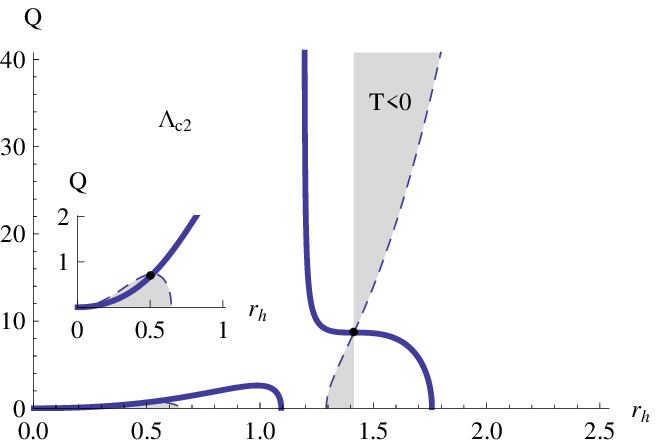}}\\
\subfigure[]{\label{PQdk-1c}
\includegraphics[width=8cm,height=6cm]{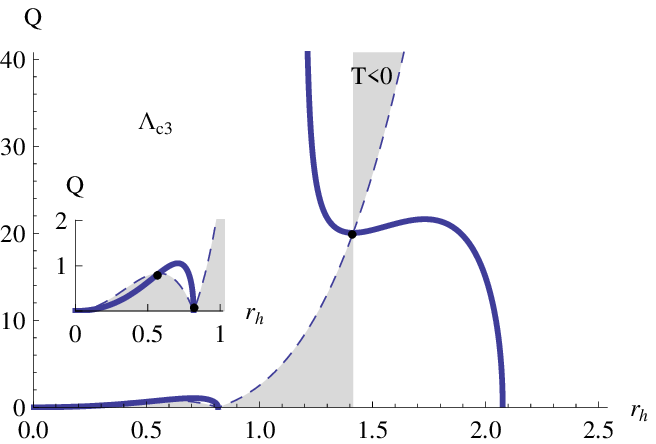}}
\subfigure[]{\label{PQdk-1d}
\includegraphics[width=8cm,height=6cm]{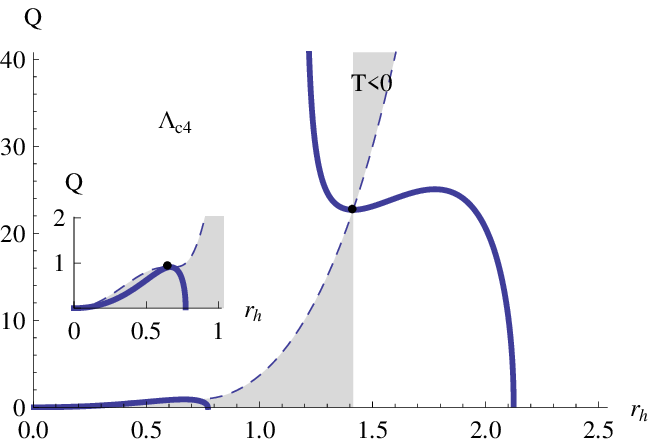}}
\caption{Four different divergent behaviors of the heat capacity $C_{Q}$ in the different ranges of the cosmological constant $\Lambda$.} \label{PQdk-1}
\end{figure*}

Now, the divergence condition (\ref{Qccondition}) for the heat capacity $C_{Q}$ reduces to
\begin{eqnarray}
 Q_{d}^{2}=\frac{2\pi r_{h}^{2d-8}
            \bigg(2\Lambda r_{h}^{4}(r_{h}^{2}-6)
             -(d-2)\left(2(d-5)-(d-9)r_{h}^{2}+(d-3)r_{h}^{4}\right)\bigg)}
             {14-5r_{h}^{2}+2d(r_{h}^{2}-2)}.
\end{eqnarray}
It is clear that the $Q_{d}^{2}$ closely depends on the cosmological constant $\Lambda$ and dimensionality $d$.

\begin{table}[h]
\begin{center}
\caption{The critical values of the cosmological constant $\Lambda$ for different behaviors of the capacity at $k=-1$.}\label{tab2}
\begin{tabular}{c c c c c c c}
  \hline
  \hline
   $d$& 5 & 6 & 7 & 8 & 9 & 10 \\
  \hline
    $\Lambda_{c1}$ & -1.50 & -2.50 & -3.75 & -5.25 & -7.00 & -9.00 \\
    $\Lambda_{c2}$ & -2.00 & -2.88 & -4.08 & -5.56 & -7.30 & -9.29 \\
    $\Lambda_{c3}$ & --- & -4.50 & -5.00 & -6.25 & -7.88 & -9.80 \\
    $\Lambda_{c4}$ & --- & -5.06 & -5.33 & -6.51 & -8.10 & -10.00 \\
\hline\hline
\end{tabular}
\end{center}
\end{table}

To study the phase structure for this case, we plot the heat capacity $C_{Q}$ in Fig. \ref{PCqk}, for the small and large values of $\Lambda$, respectively. From the figure, we see that the heat capacity behaves very different for the small and large values of $\Lambda$. And we should note that the extremal condition of the black hole is not included within. After considering that condition, four critical cosmological constants, listed in Table \ref{tab2}, are obtained. Among the different ranges of cosmological constant, the divergent behavior of the heat capacity is different, and the behaviors of these divergent points are given in Fig. \ref{PQdk-1}, for the four critical cosmological constants. In these figures, the ranges of shadow are for $T<0$; thus we do not consider these ranges. From Fig. \ref{PQdk-1}, one can easily get the change of the number of divergent points for the heat capacity as $Q$ increases. It is also clear that when $\Lambda_{c2}>\Lambda>\Lambda_{c3}$, the nonextremal black hole has a more richer phase structure.

As discussed above, the behavior of the heat capacity in each range of $\Lambda$ is almost the same for different $d$, and the phase structure is more richer if $\Lambda\in(\Lambda_{c2}, \Lambda_{c3})$. So, in order to well study the phase structure, we here set $d=6$ and $\Lambda=-4$. Then the divergence condition (\ref{Qccondition}) reduces to
\begin{eqnarray}
 Q_{d}^{2}=\frac{2\pi r_{h}^{4}
            \bigg(-20 r_{h}^{4}(r_{h}^{2}-6)
             -4\left(2+3r_{h}^{2}+3r_{h}^{4}\right)\bigg)}
             {14-5r_{h}^{2}+12(r_{h}^{2}-2)}.
\end{eqnarray}
\begin{figure*}
\centerline{\subfigure[]{\label{PQdk-14a}
\includegraphics[width=8cm,height=6cm]{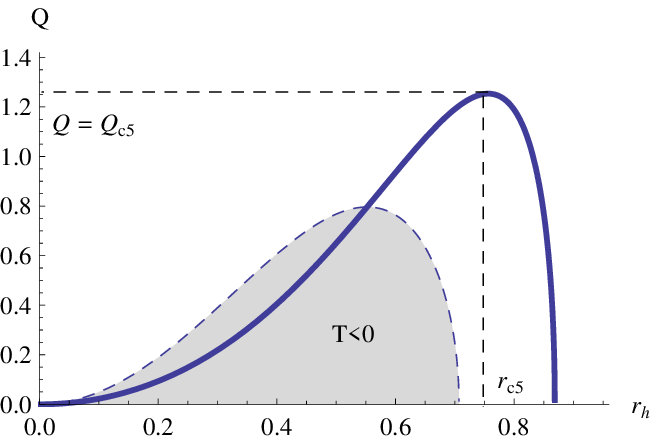}}
\subfigure[]{\label{PQdk-14b}
\includegraphics[width=8cm,height=6cm]{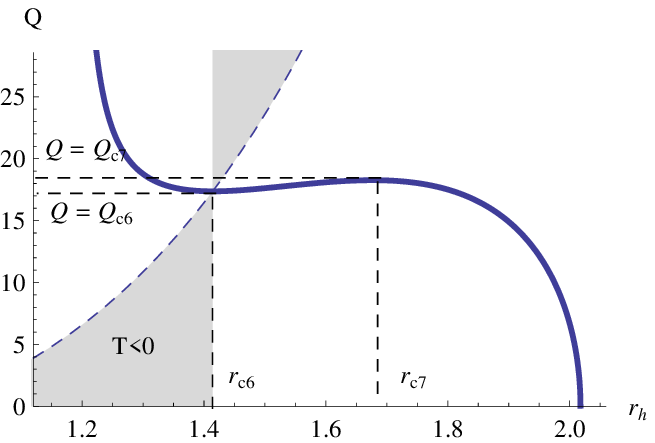}}}
\caption{The divergent behavior of $C_{Q}$ for $k=-1$, $d=6$, and $\Lambda=-4$. (a) for the small charge case, and (b) for the large charge case.} \label{PQdk-14}
\end{figure*}
The divergent behavior of the heat capacity is plotted in Fig. \ref{PQdk-14} for small charge and large charge. For the small charge $Q$, there exists a local maximum located at $r_{c5}$. For the large charge $Q$, there are one local maximum and one minimum located at $r_{c7}$ and $r_{c6}$, respectively. Here, we should point out that the critical radius $r_{c6}=\sqrt{2}$. And the point $(r_{c6}, Q_{c6})$ corresponds to a nonextremal black hole with a positive finite temperature. Among these critical points, two relations holds, i.e., $Q_{c5}<Q_{c6}<Q_{c7}$ and $r_{c5}<r_{c6}<r_{c7}$. For any other value of $d$, they share the same property. The critical values of these thermodynamic quantities are listed in Table \ref{Tab3} for different critical points.

\begin{table}[h]
\begin{center}
\caption{The critical values of $r_{c}$, $Q_{c}$, $\Phi_{c}$, $T_{c}$ at these critical points located at $r_{c5}$, $r_{c6}$, and $r_{c7}$. And the extremal black hole radius $r_{e}$.}\label{Tab3}
\begin{tabular}{cccccc}
  \hline
  \hline
   $r_{c}$ & $Q_{c}$ & $\Phi_{c}$ & $T_{c}$ & $r_{e}$ \\\hline
   1.6765 & 18.2666 & 0.0082 & 0.3915 & 1.4227\\
   1.4142 & 17.3664 & 0.0130 & 0.4538 & 1.4142\\
   0.7565 & 1.2535  & 0.0061 & 0.0186 & 1.0253\\
\hline\hline
\end{tabular}
\end{center}
\end{table}

Here, some notes on the $Q$-$\Phi$ criticality is necessary. It is clear that this case is similar as $k=1$, $d=6$, and $\Lambda=-0.49$, which have two local maxima and one local minimum. And after some tedious calculations, we find the $Q$-$\Phi$ diagram near the local maxima indeed shares the same shape as shown in Fig. \ref{PQphit}, only with different critical values of the thermodynamic quantities. However, for the critical point located at $r_{c6}$, there is no such $Q$-$\Phi$ criticality, which may be mainly due to the fact that there is no closed loop surrounding the critical point when the range of $T<0$ is excluded. And more research about this point should be carried out in the future. It is also worth noting that these two critical points located at the maxima have different values of temperature, so they are on the different isotherm of a GB-AdS black hole.

\section{Critical exponents and scaling laws}
\label{exponent}

From above discussion, we find that the $Q$-$\Phi$ criticality of a GB-AdS black hole near the critical point at the critical isotherm has an analogy as the $P$-$V$ criticality of a van der Waals liquid gas system. So we first give a brief introduction to the critical exponents for a van der Waals liquid gas system and then apply it to the charged GB-AdS black hole.

Near the critical point at the critical isotherm, the critical behavior of a van der Waals liquid gas system can be described with the following critical exponents \cite{Reichl}:
\begin{center}
\begin{enumerate}
   \item[(1)] $C_{V}\sim (-\frac{T-T_{c}}{T_{c}})^{-\alpha'}$ for $T<T_{c}$,\\
       \quad\; $\sim (\frac{T-T_{c}}{T_{c}})^{-\alpha}$ \quad for
         $T>T_{c}$,
   \item[(2)] $\frac{V_{g}-V_{l}}{V_{c}}\sim (-\frac{T-T_{c}}{T_{c}})^{\beta}$,
   \item[(3)] $\kappa_{T}\sim (-\frac{T-T_{c}}{T_{c}})^{-\gamma'}$ for $T<T_{c}$,\\
      \quad\; $\sim (\frac{T-T_{c}}{T_{c}})^{-\gamma}$ \quad for $T>T_{c}$,
   \item[(4)] $P-P_{c}\sim (V-V_{c})^{\delta}$.
\end{enumerate}
\end{center}
Here, the index $c$ denotes the critical points of the van der Waals liquid gas system, and $g$ for the gas phase and $l$ for the liquid phase. For this system, the critical exponents have the following values: $\alpha=\alpha'=0$, $\beta=1/2$, $\gamma'=\gamma=1$, and $\delta=3$.

Now, let us turn to study these critical exponents for the GB-AdS black hole. In order to examine the neighborhood of the critical point, we introduce the expansion parameters
\begin{eqnarray}
 \epsilon_{i}=\left(\frac{T}{T_{ci}}\right)-1,\;\;
 \omega_{i}=\left(\frac{\Phi}{\Phi_{ci}}\right)-1,\;\;
 \pi_{i}=\left(\frac{Q}{Q_{ci}}\right)-1.\label{small}
\end{eqnarray}
The index $i$ denotes different critical points shown in Sec. \ref{Critical}. Note that the critical point located at $r_{c6}$ is a special one, so we will not consider it here. Using (\ref{tempature}) and (\ref{phi}), we have
\begin{eqnarray}
 T=\frac{-Q^{2}+2\pi \Phi_{0}^{2d-8}
            \bigg((d-2)k((d-3)\Phi_{0}^{2}+(d-5)k\alpha)
                  -2\Lambda \Phi_{0}^{4}\bigg)}
            {8\pi^{2}(d-2)\Phi_{0}^{2d-7}(2k\alpha+\Phi_{0}^{2})}.\label{tqphi}
\end{eqnarray}
Then substituting (\ref{small}) into (\ref{tqphi}), we get
\begin{eqnarray}
 \pi_{i}=a_{00}&+&a_{10}\epsilon_{i}+a_{01}\omega_{i}+a_{11}\epsilon_{i}\omega_{i}
          +a_{20}\epsilon_{i}^{2}
     +a_{02}\omega_{i}^{2}+a_{21}\epsilon_{i}^{2}\omega_{i}\nonumber\\
    &+&a_{12}\epsilon_{i}\omega_{i}^{2}
     +a_{30}\epsilon_{i}^{3}+a_{03}\omega_{i}^{3}+...,\label{pi}
\end{eqnarray}
where $a_{mn}=\frac{1}{m!n!}\frac{\partial^{(m+n)}\pi_{i}}{\partial\epsilon_{i}^{m}\partial \omega_{i}^{n}}$. With detailed calculation, we always have
\begin{eqnarray}
 a_{00}=0,\;a_{01}=0,\;a_{02}=0.
\end{eqnarray}
This result holds for any values of $k$, $\Lambda$, and $d$. And other coefficients $a_{mn}$ have different values for different critical points. For an example, the expansion coefficients for the critical point located at $r_{c5}$ are $a_{10}=0.6619$, $a_{11}=-0.9263$, $a_{20}=0.3941$, $a_{21}=-1.8015$, $a_{12}=2.2493$, $a_{30}=0.3273$, and $a_{03}=-0.6147$.

In the neighborhood of the critical points, we have (\ref{pi}). The values of $\omega_{i}$ on either side of the coexistence curve can be found from the conditions that along the isotherm,
\begin{eqnarray}
 &&Q(\Phi_{ia})=Q(\Phi_{ib}), \label{condition1}\\
 &&\int_{\Phi_{ia}}^{\Phi_{ib}}\Phi_{i} dQ=0.\label{condition2}
\end{eqnarray}
This is just the Maxwell equal-area law. And the indices $a$ and $b$ represent two different phases of the thermodynamics system. From the first condition (\ref{condition1}), we derive
\begin{eqnarray}
 a_{11}\epsilon_{i}(\tilde{\omega}_{ib}+\tilde{\omega}_{ia})
 +a_{21}\epsilon_{i}^{2}(\tilde{\omega}_{ib}+\tilde{\omega}_{ia})
 +a_{12}\epsilon_{i}(\tilde{\omega}_{ib}^{2}-\tilde{\omega}_{ia}^{2})
 +a_{03}(\tilde{\omega}_{ib}^{3}+\tilde{\omega}_{ia}^{3})
 +...=0,\label{aaa}
\end{eqnarray}
where we have denoted $\tilde{\omega}_{ia}=-\omega_{ia}$ and $\tilde{\omega}_{ib}=\omega_{b}$ for different phases. The second condition (\ref{condition2}) reduces to
\begin{eqnarray}
 a_{11}\epsilon_{i}(\tilde{\omega}_{ib}+\tilde{\omega}_{ia})
 +a_{21}\epsilon_{i}^{2}(\tilde{\omega}_{ib}+\tilde{\omega}_{ia})
 +\frac{1}{2}(a_{11}+2a_{12})\epsilon_{i}(\tilde{\omega}_{ib}^{2}-\tilde{\omega}_{ia}^{2})
 +a_{03}(\tilde{\omega}_{ib}^{3}+\tilde{\omega}_{ia}^{3})
 +...=0.\nonumber\\\label{bbb}
\end{eqnarray}
Requiring that (\ref{aaa}) is consistent with (\ref{bbb}), we have
\begin{eqnarray}
 \tilde{\omega}_{ia}^{2}=\tilde{\omega}_{ib}^{2}
  =-\frac{1}{a_{03}}(a_{11}\epsilon_{i}+a_{21}\epsilon_{i}^{2}).
\end{eqnarray}
Therefore
\begin{eqnarray}
 \tilde{\omega}_{ia}=\tilde{\omega}_{ib}
  \sim \sqrt{-\frac{a_{11}}{a_{03}}\epsilon_{i}}\sim \epsilon_{i}^{1/2},
\end{eqnarray}
which implies that the degree of the coexistence curve
\begin{eqnarray}
 \beta=1/2.
\end{eqnarray}
Let $\epsilon_{i}=0$, (\ref{pi}) will be of the form
\begin{eqnarray}
 \pi_{i}=a_{03}\omega_{i}^{3}+\cdot\cdot\cdot\;
\end{eqnarray}
Thus we get the degree of the critical isotherm
\begin{eqnarray}
 \delta=3.
\end{eqnarray}
With (\ref{tempature}), (\ref{phi}), and (\ref{entropy}), the heat capacity at fixed $\Phi$ is
\begin{eqnarray}
 C_{\Phi}=T\bigg(\frac{\partial S}{\partial T}\bigg)_{\Phi}
         =T\bigg(\frac{\partial S}{\partial Q}\bigg)_{\Phi}
           \bigg(\frac{\partial T}{\partial Q}\bigg)_{\Phi}^{-1}
         =-\frac{(d-2)(r_{h}^{2}+2k)^{2}(Q^{2}-Q_{e}^{2})}
          {4r_{h}^{12-d}(r_{h}^{2}-2k)(Q^{2}-\Theta)}.
\end{eqnarray}
From above equation, we see that $C_{\Phi}$ has no singular behavior at these critical points, while it is singular at $Q^{2}=\Theta$. Thus the heat capacity exponent
\begin{eqnarray}
 \alpha=\alpha'=0.
\end{eqnarray}
The isothermal compressibility can be expressed as
\begin{eqnarray}
 K_{T}^{-1}\propto \left(\frac{\partial\pi_{i}}{\partial\omega_{i}}\right)_{\epsilon}
           =a_{11}\epsilon_{i}+a_{21}\epsilon_{i}^{2}+2a_{12}\epsilon_{i}\omega_{i}
             +3a_{03}\omega_{i}^{2}.
\end{eqnarray}
When $T<T_{ci}$ (or $T>T_{c2}$), one approaches the critical point along the critical isochore. Then set $\omega_{i}=0$, we get
\begin{eqnarray}
 K_{T}^{-1}\propto a_{11}\epsilon_{i}.
\end{eqnarray}
Thus the isothermal compressibility exponent
\begin{eqnarray}
 \gamma'=1.
\end{eqnarray}
When $T>T_{ci}$ (or $T<T_{c2}$), one approaches the critical point along the coexistence curve. Then set $\omega_{i}=\sqrt{-\frac{a_{11}}{a_{03}}\epsilon_{i}}$, and we have
\begin{eqnarray}
 K_{T}^{-1}\propto -2a_{11}\epsilon.
\end{eqnarray}
And the isothermal compressibility exponent
\begin{eqnarray}
 \gamma=1.
\end{eqnarray}
With these results, these critical exponents are found to satisfy the following relations, known as the thermodynamic scaling laws,
\begin{eqnarray}
 &&\alpha+2\beta+\gamma=2, \;\;\quad\quad\quad\quad\quad\alpha+\beta(1+\delta)=2,\\
 &&\gamma(1+\delta)=(2-\alpha)(\delta-1),\;\quad\gamma=\beta(\delta-1).
\end{eqnarray}
When sufficiently close to the critical points $r_{ci}$, the Gibbs free energy for a GB-AdS black hole may expand as
\begin{eqnarray}
 G(T, \Phi)=&\bigg[G\bigg]_{ci}+\bigg[\bigg(\frac{\partial G}
                {\partial T}\bigg)_{\Phi}\bigg]_{ci}(T-T_{ci})
            +\bigg[\bigg(\frac{\partial G}
              {\partial \Phi}\bigg)_{T}\bigg]_{ci}(\Phi-\Phi_{ci})\nonumber\\
           &+\frac{1}{2}
             \bigg[\bigg(\frac{\partial^{2}G}
               {\partial T^{2}}\bigg)_{\Phi}\bigg]_{ci}(T-T_{ci})^{2}
           +\frac{1}{2}\bigg[\bigg(\frac{\partial^{2}G}
            {\partial \Phi^{2}}\bigg)_{T}\bigg]_{ci}(\Phi-\Phi_{ci})^{2}+...\;.
\end{eqnarray}
According to the thermodynamics law of this black hole $dG=-SdT-Qd\Phi$, we have
\begin{eqnarray}
 S&=&-\bigg(\frac{\partial G}{\partial T}\bigg)_{\Phi},\;\;\quad
 Q=-\bigg(\frac{\partial G}{\partial \Phi}\bigg)_{T},\\
 C_{\Phi}&=&-T\bigg(\frac{\partial^{2}G}{\partial T^{2}}\bigg)_{\Phi},\;
 K_{T}^{-1}=\bigg(\frac{\partial^{2}G}{\partial \Phi^{2}}\bigg)_{T}.
\end{eqnarray}
Since the entropy and charge are regular for any values of these parameters, the singular part of Gibbs free energy can be expressed as
\begin{eqnarray}
 G_{s}\sim -\frac{1}{2T_{c}}C_{\Phi}\big|_{ci}(T-T_{ci})^{2}
       -\frac{1}{2}K_{T}^{-1}\big|_{ci}(\Phi-\Phi_{ci})^{2}.
\end{eqnarray}
Express it with the small parameters $\epsilon$ and $\pi$, we derive
\begin{eqnarray}
 G_{s}\sim -\frac{1}{2}C_{\Phi}\big|_{ci}\epsilon^{2}
           -\frac{1}{2}\Phi_{ci}^{2}\pi^{4/3}.\label{gg}
\end{eqnarray}
From (\ref{gg}), we find
\begin{eqnarray}
 G_{s}(\lambda^{\varphi}\epsilon, \lambda^{\psi}\omega)
   =\lambda G_{s}(\epsilon, \omega),
\end{eqnarray}
with $\varphi=1/2$, $\psi=3/4$, and $\lambda$ a real number. This strongly suggests that the generalized homogeneous function hypothesis holds for this charged GB-AdS black hole, which means that, close to these critical points, the singular part of the Gibbs free energy is a generalized homogeneous function. Especially, similar to the case of a van der Waals liquid gas system, the critical exponents derived above can be expressed in terms of $\varphi$, $\psi$ as \cite{Reichl,Stanley}
\begin{eqnarray}
 \alpha&=&2-\frac{1}{\varphi},\;\;\;\beta=\frac{1-\psi}{\varphi},\\
 \gamma&=&\frac{2\psi-1}{\varphi},\;\delta=\frac{\psi}{1-\psi}.
\end{eqnarray}
Now, we see that the critical exponents and scaling laws of the black hole near these critical points are the same as that of a van der Waals liquid gas system.

\section{State space scalar curvature}
\label{curvature}

In above sections, we study the phase structure and critical phenomena for a charged GB-AdS black hole. These results strongly imply that this black hole thermodynamics system is similar to the van der Waals liquid gas system. However, the microscopic degree of freedom of the van der Waals liquid gas system is clear, while for the black hole system, it is still unclear. In the context of thermodynamical fluctuation theory, Ruppeiner \cite{Ruppeiner} has argued that the Riemannian geometry could give insight into the underlying statistical mechanical system. The divergent behavior of Ruppeiner curvature is related to the phase transition and the sign of it represents the attraction or repulsion among the microscopic degree of freedom. For the van der Waals liquid gas system, the curvature is negative and goes to negative infinity at the critical point of the phase transition. In this section, we will explore the Ruppeiner geometry for the charged GB-AdS black hole.

The Ruppeiner metric is defined as \cite{Ruppeiner}
\begin{eqnarray}
 ds^{2}_{R}=-\frac{\partial^{2}S}{\partial x^{i}\partial x^{j}}dx_{i}dx_{j},
\end{eqnarray}
with $x_{i}$ the extensive variables of the system. On the other hand, there exists another metric proposed by Weinhold, which is defined in the following way \cite{Weinhold}:
\begin{eqnarray}
 ds^{2}_{W}=-\frac{\partial^{2}U}{\partial x^{i}\partial x^{j}}dx_{i}dx_{j},
\end{eqnarray}
where $U$ is the internal energy of the system. These two geometries are found to be conformally related with each other,
\begin{eqnarray}
 ds^{2}_{R}=\frac{1}{T}ds^{2}_{W}.
\end{eqnarray}
For the GB-AdS black hole, we chose $U=M-Q \Phi$, and $x_{1}=S$, $x_{2}=\Phi$. With Eqs. (\ref{mass}), (\ref{tempature}), (\ref{phi}), and (\ref{entropy}), the Ruppeiner metric is calculated in terms of $Q$ and $r_{h}$
\begin{eqnarray}
 g_{SS}&=&\frac{Q^2 \left(r_{h}^2+2\right)+8
   \pi  \left(2 r_{h}^6-9r_{h}^4+3 r_{h}^2+2\right)
   r_{h}^4}{r_{h}^2 \left(r_{h}^2-2\right)^2
   \left(Q^2-8\pi r_{h}^4
   \left(2r_{h}^4-3r_{h}^2+1\right)\right)},\nonumber\\
 g_{S\Phi}&=&g_{\Phi S}=-\frac{96 \pi ^2 Q r_{h}^3}{Q^2-8\pi
   r_{h}^4 \left(2 r_{h}^4-3
   r_{h}^2+1\right)},\\
 g_{\Phi\Phi}&=&-\frac{1536 \pi ^4 r_{h}^8
   \left(r_{h}^2-2\right)}{Q^2-8 \pi
   r_{h}^4 \left(2 r_{h}^4-3
   r_{h}^2+1\right)}.\nonumber\\
\end{eqnarray}
The definitions of the Christoffel symbols and Riemann curvature tensor are \begin{eqnarray}
  \Gamma^{\lambda}_{\;\;\mu\nu}
     &=&\frac{1}{2}g^{\lambda\tau}(g_{\nu\tau,\mu}+g_{\mu\tau,\nu}-g_{\mu\nu,\tau}),\\ R^{\mu}_{\;\;\sigma\nu\tau}
    &=&\Gamma^{\mu}_{\;\;\sigma\nu,\tau}-\Gamma^{\mu}_{\;\;\sigma\tau,\nu}
                         +\Gamma^{\mu}_{\;\;\lambda\tau}\Gamma^{\lambda}_{\;\;\sigma\nu}
                         -\Gamma^{\mu}_{\;\;\lambda\nu}\Gamma^{\lambda}_{\;\;\sigma\tau}.
\end{eqnarray}
Then the Ruppeiner curvature has the following form
\begin{eqnarray}
 \mathcal{R}\sim \frac{1}{(Q^{2}-Q_{d}^{2})^{2}}T^{-1}
      \sim C_{Q}^{2}T^{-1}.
\end{eqnarray}
From above equation, we clear see that, Ruppeiner curvature shares the same divergent behavior as that of the heat capacity $C_{Q}$ (\ref{heatcapacity}). Besides, $\mathcal{R}$ also diverges for an extremal black hole, where $T=0$. These imply that the phase transition is included in the Ruppeiner curvature $\mathcal{R}$.

\begin{figure*}
\subfigure[]{\label{PRra}
\includegraphics[width=8cm,height=6cm]{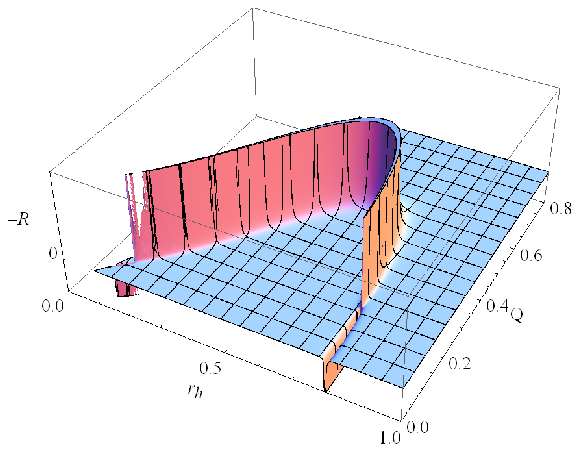}}
\subfigure[]{\label{PRrb}
\includegraphics[width=8cm,height=6cm]{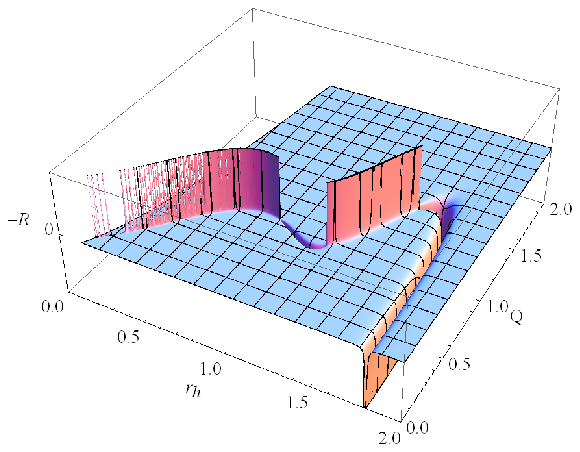}}\\
\centerline{\subfigure[]{\label{PRrc}
\includegraphics[width=8cm,height=6cm]{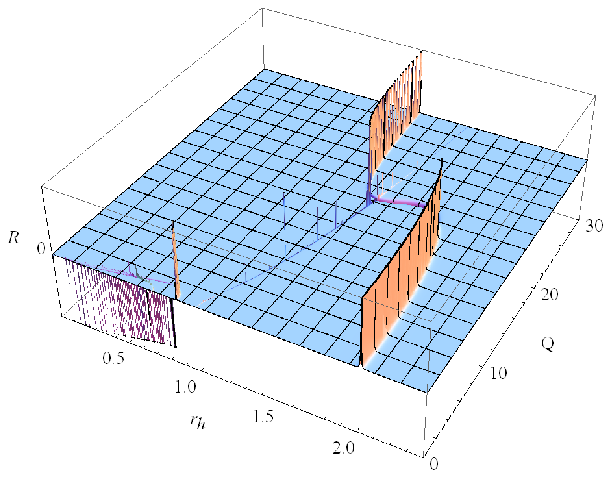}}}
\caption{Behavior of the Ruppeiner curvature at $d=6$. (a) for $k=1$, $\Lambda=-1$, (b) for $k=1$, $\Lambda=-0.49$, (c) for $k=-1$, $\Lambda=-4$.} \label{PRr}
\end{figure*}

We plot the Ruppeiner curvature in Fig. \ref{PRr} for different $r_{h}$ and $Q$. It is obvious that the divergent behavior of the Ruppeiner curvature is exactly consistent with the heat capacity described in Fig. \ref{PCqk1} and  \ref{PCqk-1b}. Note that the divergent behavior located at the extremal black hole has been suppressed in these figures. Moreover, we can see that in different ranges of $r_{h}$ and $Q$, the Ruppeiner curvature has a different sign, which represents the different interactions among the microscopic degrees of freedom. It is worth noting that along the divergent curve, the Ruppeiner curvature also has a different sign. The range of the positive sign is similar to the ideal gas of fermions, whose curvature is found to be positive and positive diverges at absolute zero \cite{Janyszek,Oshima}, which can be understood with the Pauli's exclusion principle forbidding two particles in the same state with unlimited repulsive force \cite{Wei}. While the range of the negative sign is similar to the ideal gas of bosons, whose curvature is negative and goes to negative infinity at absolute zero, it appeared as Bose-Einstein condensation \cite{Janyszek}. This property is unlike the van der Waals liquid gas system, whose Ruppeiner curvature is nonpositive for different range of the parameters, which means repulsive force cannot exist in such thermodynamics system. While for the GB-AdS black hole, it has a multiple phenomena, both the repulsive force and attractive force are allowed, depending on the values of these parameters. So, it possesses both the property as the ideal gas of fermions and ideal gas of bosons.

\section{Discussions and conclusions}
\label{Conclusion}

In this paper, we have studied various aspects of the phase structure, criticality, scaling behavior, and state space scalar curvature of $d$-dimensional charged topological GB-AdS black holes. Several observations leading to important new insights to these black holes have been obtained in this context. We find that the phase structure in the canonical ensemble significantly depends on the parameter $k$, dimensionality $d$, and cosmological constant $\Lambda$.

When $k=1$, there exists no divergent behavior for $d=5$, which indicates no phase transition takes place. For $d=6$, in different ranges of the cosmological constant $\Lambda$, there are different critical behaviors. For $\Lambda\in(-\infty, -0.5)\cup(-0.4725, 0)$, there exists one local maximum, corresponded to a critical point, along the divergent curve of the heat capacity. While for $\Lambda\in(-0.5, -0.4725)$, three critical points appear. For $d\geq 7$, its critical behavior is similar to that of $d=6$ with $\Lambda\in(-\infty, -0.5)\cup(-0.4725, 0)$.

When $k=0$, the heat capacity $C_{Q}$ is always regular for any values of the dimensionality $d$ and cosmological constant $\Lambda$. Thus, phase transition does not take place in this case.

When $k=-1$, it becomes complicated. We show that the heat capacity of a nonextremal black hole has four divergent behaviors for different values of $\Lambda$. When $\Lambda$ is in a proper range, there will be three critical points. And this result is found to be held for any value of the dimensionality $d$.

Further, we plot the $Q$-$\Phi$ diagram near these critical points. The result indicates that the $Q$-$\Phi$ diagram for the critical points located at the maxima and minima is different from each other. These diagrams also show a novel liquid-gas like first-order phase transition in the canonical ensemble, culminating at a second-order critical point. In the neighborhood of the critical points, we examine the critical exponents for the charged GB-AdS black hole. The result shows that the values of these critical exponents are the same as the RN-AdS black hole with spherically symmetric case, which posses a van der Waals-like critical behavior. However, for the five-dimensional spherical GB black holes and $d$-dimensional planar GB black holes, there is no such critical behavior. But there may be another type of phase transition associated with a scalar hair in the planar case \cite{Hertog}.

Moreover, using the state space scalar curvature, we study the phase transition and explore the possible interaction among the microscopic degrees of freedom. The Ruppeiner curvature constructed in this paper shares the same divergent behavior as the heat capacity $C_{Q}$, which implies that the information of phase transition is contained in the curvature. Different from the van der Waals system, where the Ruppeiner curvature is always negative, the curvature for the GB-AdS black hole can be negative or positive, possibly produced by the attractive force or repulsive force. And along the divergent curve, it changes the sign, which indicates it has multistructures of the microscopic degrees of freedom. Although we still do not understand the microscopic structure of the horizon of a black hole, the phase structure and the state space scalar curvature may provide a useful probe to study the inner degrees of freedom of the horizon.

\section*{Acknowledgements}
This work was supported by the Program for New Century Excellent Talents in University, the Huo Ying-Dong Education Foundation of the Chinese Ministry of Education (No. 121106), the Fundamental Research Funds for the Central Universities (No. lzujbky-2012-k30), the National Natural Science Foundation of China (No. 11075065), and the National Natural Science Foundation of China (No. 11205074).

\end{document}